\documentclass[%
reprint,
superscriptaddress,
showpacs,preprintnumbers,
amsmath,amssymb,
aps,
]{revtex4-1}

\usepackage{color}
\usepackage{graphicx}
\usepackage{dcolumn}
\usepackage{bm}
\DeclareGraphicsExtensions{.png, .bb}
\begin{document}

\preprint{}

\title{
High mobility charge transport in a multicarrier altermagnet CrSb
}

\author{Takahiro Urata}
\thanks{urata@mp.pse.nagoya-u.ac.jp}
\affiliation{Department of Materials Physics, Nagoya University, Chikusa-ku, Nagoya 464-8603, Japan}

\author{Wataru Hattori}
\affiliation{Department of Materials Physics, Nagoya University, Chikusa-ku, Nagoya 464-8603, Japan}

\author{Hiroshi Ikuta}
\affiliation{Department of Materials Physics, Nagoya University, Chikusa-ku, Nagoya 464-8603, Japan}
\affiliation{Research Center for Crystalline Materials Engineering, Nagoya University, Chikusa-ku, Nagoya 464-8603, Japan}

\date{\today}

\begin{abstract}
A newly identified magnetic phase called altermagnet is being actively studied because of its unprecedented spin-dependent phenomena.
Among the candidate materials, CrSb has a particularly high ordering temperature and a large spin-splitting energy, but its transport properties have remained unexplored.
In this study, we report the magnetotransport properties of CrSb measured on single crystals.
We found that the Hall resistivity shows a nonlinear dependence on the magnetic field at low temperatures.
From symmetry-based considerations, however, this behavior can not be attributed to an anomalous Hall effect, but to a multicarrier effect. 
A multicarrier fitting to the in-plane conductivity tensor revealed the presence of carriers with high mobility in CrSb, which is an advantage for efficient spin current generation.
\end{abstract}

\maketitle

\section{Introduction}
By reclassifying antiferromagnetic materials using spin groups, a novel state called altermagnet has been proposed recently \cite{Smejkal2022a,Smejkal2022b,Mazin2022}.
In this state, the time-reversal ($\cal{T}$), as well as $\cal{PT}$ and $t\cal{T}$ symmetries, are broken, where $\cal{P}$ and $t$ represent a space inversion and a lattice translation, respectively.
Owing to these symmetry breaking, the electronic bands are spin-split akin to ferromagnets even in the absence of spin-orbit interaction, although the spins are compensated and the macroscopic magnetization is zero.
This has indeed been observed in several altermagnets by angle-resolved photoemission spectroscopy \cite{fedchenko_observation_2024,krempasky_altermagnetic_2024,lee_broken_2024,osumi_observation_2024,reimers_direct_2024}.
Furthermore, ferromagnetic-like phenomena, such as the anomalous Hall effect (AHE) \cite{feng_anomalous_2022,gonzalez_spontaneous_2023,tschirner_saturation_2023,wang_emergent_2023}, the spin current generation \cite{bose_tilted_2022,bai_observation_2022}, and the piezomagnetic effect \cite{aoyama_piezomagnetic_2023} have been reported.
These interesting discoveries have brought large attention to altermagnets as novel spintronic materials.

Among the large number of altermagnets that were theoretically predicted \cite{Smejkal2022b}, RuO$_2$ \cite{fedchenko_observation_2024,feng_anomalous_2022,tschirner_saturation_2023,wang_emergent_2023,bose_tilted_2022,bai_observation_2022} and MnTe \cite{krempasky_altermagnetic_2024,lee_broken_2024,osumi_observation_2024,gonzalez_spontaneous_2023,aoyama_piezomagnetic_2023} are the most extensively studied.
Compared to them, CrSb has received less attention, despite its notably high ordering temperature of $\sim$700 K \cite{hirone_magnetic_1956} and significantly large spin-splitting energy \cite{Smejkal2022b,reimers_direct_2024}.
This compound crystallizes in a hexagonal NiAs-type structure \cite{hirone_magnetic_1956}, and its magnetic structure, as determined by neutron diffraction measurements, exhibits a collinear alignment of antiparallel spins along the $c$ axis \cite{snow_neutron_1952}.
The crystal and magnetic structures are shown in Fig. \ref{XRD}(a).
The magnetic sublattices are connected by a screw operation $6_3$ along the $c$ axis or by a mirror operation perpendicular to it.
This characteristic classifies CrSb as a so-called bulk $g$-wave altermagnet \cite{Smejkal2022a}, where the spin-splitting bands alter their polarity six times around the $k_z$ axis. 
Given the complex nature of the spin band splitting, it is reasonable to expect that unique behavior would manifest itself in the transport properties.
However, a detailed study on the transport properties using single crystals has been lacking.

In this study, we report on the magnetotransport properties of CrSb single crystals.
We observed that the Hall resistivity depended nonlinearly on the magnetic field ($H$) at low temperatures, the origin of which is likely a multicarrier effect.
The multicarrier analysis indicated the presence of four types of carriers in this system, two of which demonstrating high mobilities exceeding $2000\ \rm{cm}^2/\rm{Vs}$.
To the best of our knowledge, these values are the highest reported among the altermagnets so far.

\section{Experimental}
Single crystals of CrSb were grown by a Sn-flux method.
Elemental Cr, Sb and Sn were weighted in a molar ratio of $1: 1: 15$, placed in an alumina crucible, and sealed in an evacuated quartz ampoule.
The ampoule was heated to 900$^{\circ}$C where it was kept for 48 h, followed by a slow cooling to 400$^{\circ}$C, and centrifuged to remove the flux.
The obtained samples were hexagonal-rod like in shape, as schematically represented in Fig. \ref{XRD}(b).
The transport properties were measured using a conventional four-terminal method with a physical property measurement system (PPMS), Quantum Design.
Hereafter, we use the tensor notation of resistivity $\rho_{ij}$ ($i,j = x, y, z$), which represents the resistivity calculated from the voltage along the $i$-direction induced by the current flowing parallel to the $j$-direction.
The measured longitudinal resistivity ($\rho_{ii}$) was symmetrized with respect to the applied magnetic field, while the Hall resistivity ($\rho_{ij}, i \ne j$) was anti-symmetrized.
The magnetization measurements were conducted with a magnetic property measurement system (MPMS), Quantum Design.
Some of the resistivity tensor components were measured using micro-sized lamellae, which were cut out from a bulk single crystal using a Ga-source focused ion beam (FIB) instrument.

\section{Results and Discussion}

\begin{figure}
    \includegraphics[width=1.0\linewidth]{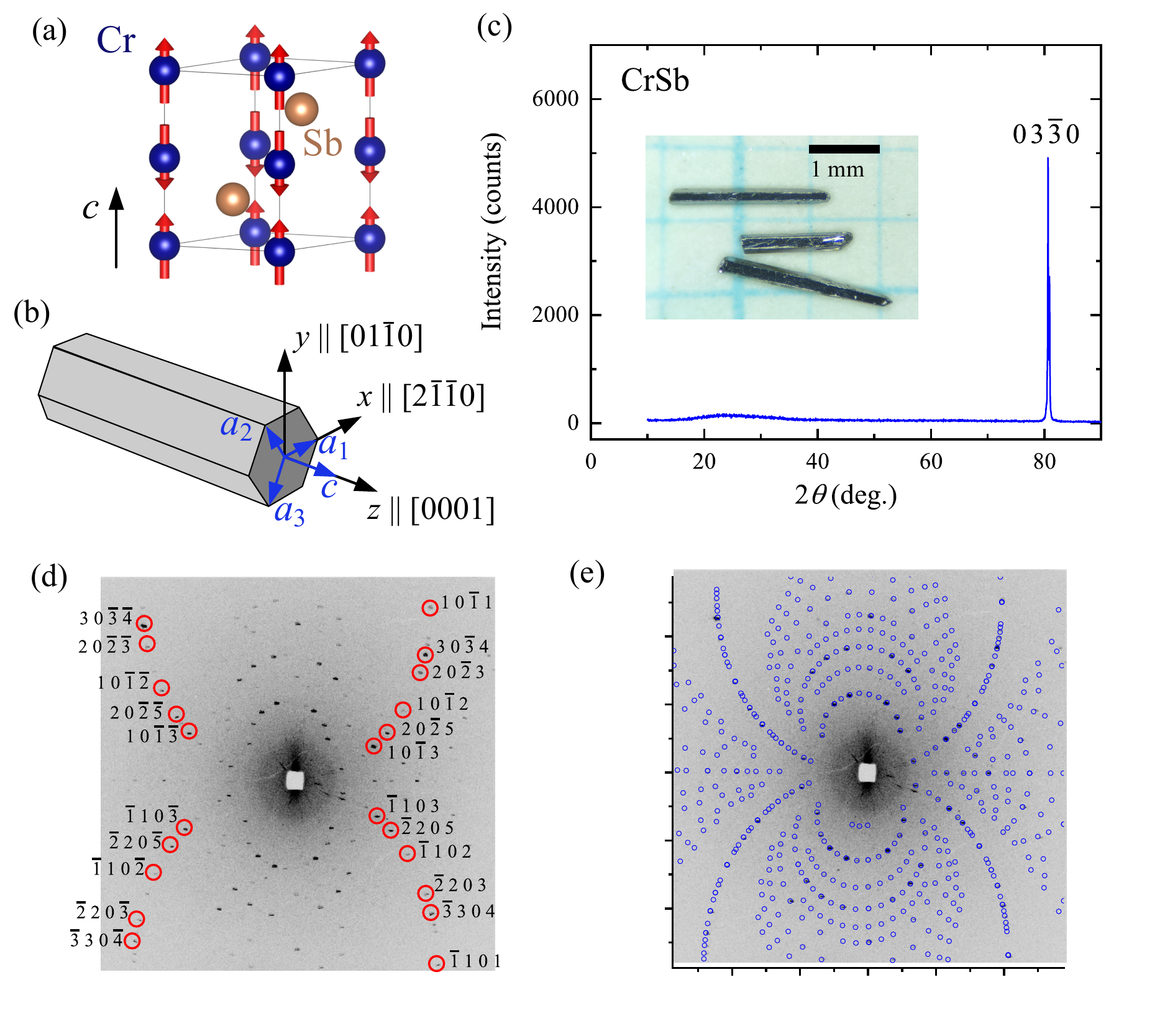}
    \caption{(Color online) (a) Crystal and magnetic structures of CrSb drawn by VESTA \cite{VESTA}.
	    (b) A schematic drawing showing the crystal shape, the crystal orientations, and the attached orthogonal axes.
	(c) Out-of-plane x-ray diffraction pattern of a CrSb single crystal.
	The inset shows a photograph of typical single crystals obtained in this work.
		(d) Transmission Laue photograph with the representative diffraction spots indexed according to the crystal structure of CrSb.
		(e) The simulated Laue pattern overlapped on the experimental one.
	        }
\label{XRD}
\end{figure}

Figure \ref{XRD}(c) shows the out-of-plane x-ray diffraction pattern obtained from the lateral facet of the crystal rod.
Only the $03\bar{3}0$ peak was observed, indicating the facet being the $(01\bar{1}0)$ plane.
Transmission Laue photographs were taken with a white light x-ray source, with the incident x-ray aligned parallel to the $[01\bar{1}0]$ direction.
As shown in Fig. \ref{XRD}(d), we observed distinct diffraction spots, and the diffraction pattern matched well with the simulation based on the crystal structure of CrSb (Fig. \ref{XRD}(e)).
These results ensure that the obtained samples are single crystals of CrSb, and the orientations of the lateral and basal facets were determined to be $\{01\bar{1}0\}$ and $\{0001\}$, respectively, as schematically illustrated in Fig. \ref{XRD}(b).
Note that the slight asymmetry seen in the Laue pattern is due to a small misalignment of the incident x-ray from the $[01\bar{1}0]$ direction.

\begin{figure*}
	\includegraphics[width=0.85\linewidth]{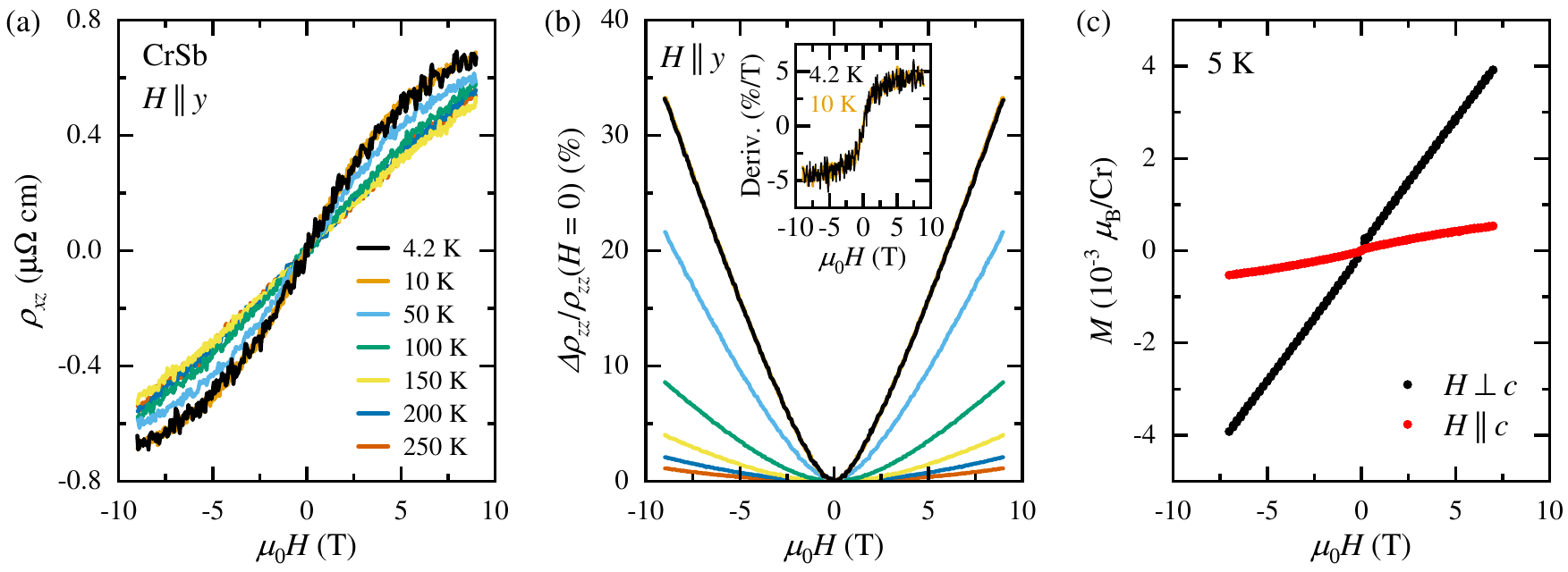}
	\caption{(Color online) (a),(b) Magnetic field dependence of Hall resistivity and magnetoresistance ratio measured at various temperatures.
	    The temperatures at which the curves in panel (b) were measured correspond to those indicated by the same color in panel (a).
	    The inset of panel (b) shows the first derivative of magnetoresistance with respect to the applied field.
    (c) Magnetic field dependence of magnetization measured with the two different field orientations.
    }
\label{MR_Hall}
\end{figure*}

Figure \ref{MR_Hall}(a) shows the magnetic field dependence of Hall resistivity $\rho_{xz}$ measured at various temperatures.
The slope is positive at all temperatures, indicating that the dominant carriers are holes.
At low temperatures, $\rho_{xz}$ exhibited a nonlinear field dependence.
Since CrSb is an altermagnet, it is tempting to attribute this behavior to the AHE.
However, we can exclude this possibility, as will be discussed later.
Figure \ref{MR_Hall}(b) shows the magnetic field dependence of magnetoresistance ratio (MR), defined as $\Delta\rho_{ii}/\rho_{ii}(H = 0) \equiv 100\% \times [\rho_{ii} - \rho_{ii}(H = 0)]/\rho_{ii}(H = 0)$.
MR was large in magnitude, exceeding 30\% at low temperatures, and showed no sign of saturation up to 9 T.
A non-saturating positive magnetoresistance can be ascribed to the coexistence of electron- and hole-type carriers \cite{Ashcroft_Mermin,ali_large_2014}.
If both types of carriers contribute equally, the magnetoresistance exhibits a $H^2$ dependence up to high magnetic fields.
However, as shown in the inset of Fig. \ref{MR_Hall}(b), the first derivative of MR with respect to the applied field is linear only at low fields.
This indicates that the compensation between electrons and holes is not complete.

Figure \ref{MR_Hall}(c) shows the magnetic field dependence of the magnetization at 5 K measured for $H \perp c$ and $H \parallel c$.
For both configurations, the magnetization depended linearly to the applied field, with no evidence of spin-flop or metamagnetic transitions.
In addition, its magnitude is small, on the order of $10^{-3} \mu_{\rm B}/{\rm Cr}$ even at 7 T, which is significantly smaller than the value of $\sim$$2.5 \mu_{\rm B}/{\rm Cr}$ estimated from a neutron scattering experiment \cite{yuan_magnetic_2020}.
These results indicate that the antiferromagnetically ordered spins are strongly coupled and can only slightly incline towards the field directions.
The magnetic susceptibility was larger when the magnetic field was applied perpendicular to the $c$ axis, which is consistent with the magnetic structure of CrSb.

From a symmetry point of view, a spontaneous AHE can only be expected if the magnetic point group (MPG) is compatible with the pseudovector comprised of the antisymmetric Hall conductivity components \cite{smejkal_crystal_2020}.
In other words, the MPG should allow a ferromagnetic state.
An example of this is the altermagnet MnTe.
The MPG of MnTe ($m'm'm$) is compatible with ferromagnetism \cite{kunitomi_neutron_1964}, and a spontaneous AHE was indeed observed in MnTe \cite{gonzalez_spontaneous_2023}. 
However, this is not the case for CrSb. 
While MnTe and CrSb share the same crystal structure, their MPGs are different. 
The MPG of CrSb ($6'/m'mm'$) is not compatible with ferromagnetism, and therefore, a spontaneous AHE can not be expected in CrSb \footnote{It is worth mentioning that an AHE may be induced in CrSb if the MPG can be changed. This was indeed demonstrated recently by Zhou \textit{et al.}, who fabricated CrSb thin films with a distorted crystal structure \protect\cite{zhou_crystal_2024}.}.

There remains, however, another possibility that could lead to an AHE, namely the application of a magnetic field $\bm{H}$ may have altered the MPG by rotating the N\'{e}el vector $\bm{L}$.
Like CrSb, the ground state MPG of the altermagnet RuO$_2$ is incompatible with ferromagnetism.
Nevertheless, the Hall resistance of a RuO$_2$ thin film exhibited a nonlinear field dependence \cite{feng_anomalous_2022,tschirner_saturation_2023}.
It is considered that the application of $\bm{H}$ perpendicular to $\bm{L}$ had caused a rotation of $\bm{L}$, changing the MPG to a one that is compatible with an AHE.
For $\bm{L}$ to rotate by the application of $\bm{H}$, the cross terms between the components of $\bm{L}$ and the macroscopic magnetization ($\bm{m}$) must have a non-zero contribution to the thermodynamic free energy \cite{dzyaloshinskii_thermodynamic_1957,dzyaloshinskii_magnetic_1958,bazhan_weak_1976}.
To satisfy this condition, $\bm{L}$ and $\bm{m}$ must have components that are the bases of the same irreducible representation (IR) of the crystal point group \cite{dzyaloshinskii_thermodynamic_1957}.
We listed all possible magnetic structures compatible with the crystal symmetry of CrSb using SARAh \cite{wills_new_2000}, but found that none of their IR met the condition.
Therefore, the possibility that $\bm{L}$ rotated by the application of $\bm{H}$ can also be excluded, and the observed nonlinear behavior of the Hall resistivity can not be attributed to the AHE.
Coupled with the absence of any indication of a field-induced magnetic transition as shown in Fig. \ref{MR_Hall}(c), the most plausible explanation would then be a multicarrier effect.
Here, it is worthwhile mentioning that, in general, even when symmetry allows an AHE, it is difficult to distinguish the origin of nonlinear Hall conductivity from multicarrier effects, and care should be taken when interpreting the data.

\begin{figure}
    \includegraphics[width=1.0\linewidth]{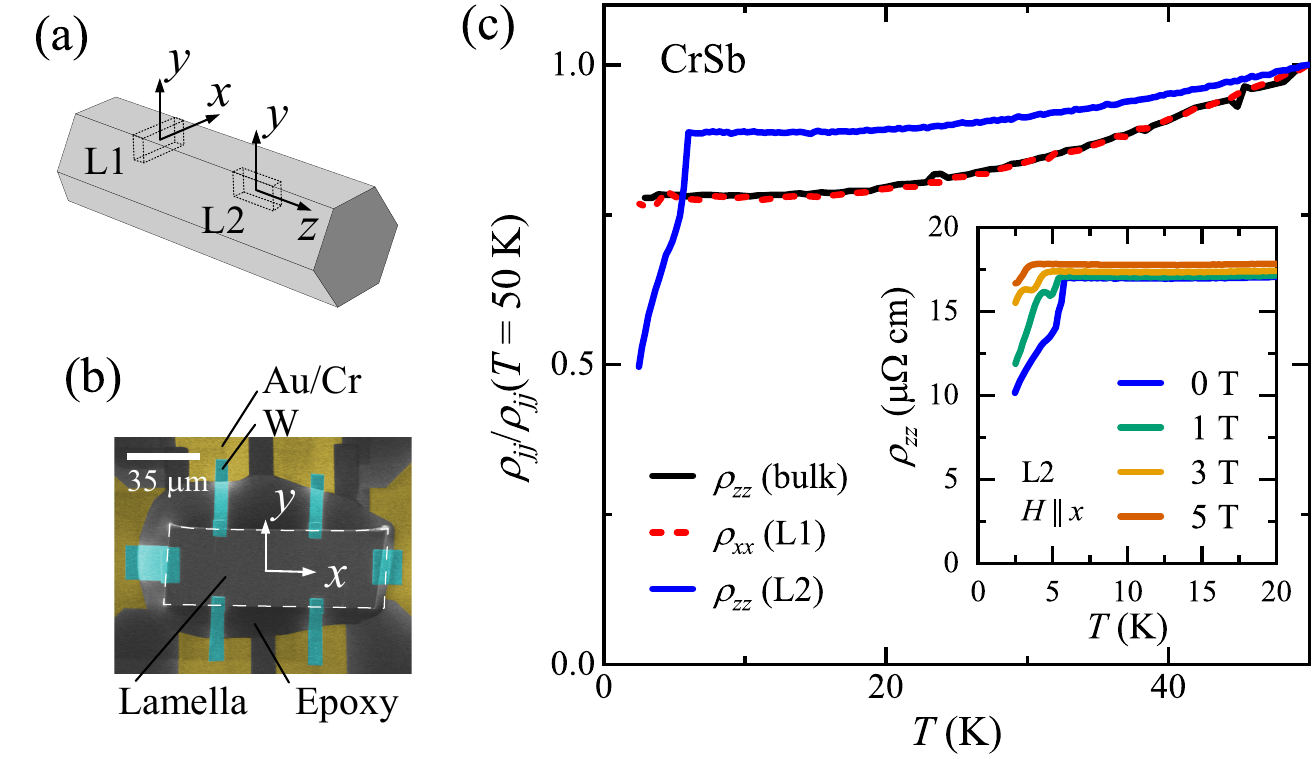}
    \caption{(Color online) (a) A schematic drawing of how the lamellae were cut out from the bulk single crystal of CrSb by FIB.
        (b) A scanning ion microscope image of the completed sample used for measuring $\rho_{xx}$ and $\rho_{yx}$.
	(c) Temperature dependence of normalized resistivity.
	The inset shows the temperature dependence of resistivity ($\rho_{zz}$) measured for sample L2 under various magnetic fields.
}
\label{FIB_RT}
\end{figure}

To gain a more comprehensive understanding of the Hall effect in CrSb, measurements using different configurations were carried out.
The Hall effect can generally be expressed with a third-rank tensor $R_{ijk}$ as $E_i = R_{ijk}J_jH_k$, where $E_i$ is the generated electric field, $J_j$ the applied current density, $H_k$ the applied magnetic field.
In standard Hall resistivity measurements, these three vectors are mutually orthogonal ($i \neq j \neq k$), allowing six different configurations.
However, considering the intrinsic symmetry of the Hall effect ($R_{ijk}=-R_{jik}$) and the MPG of CrSb ($R_{yzx}= R_{zxy}$) \cite{Birss}, there are only two independent components, namely, $R_{xyz}$ and $R_{yzx}$.
Consequently, the Hall resistivity tensor components for the three field directions are expressed by the following equations.
\begin{align}
	\rho_{xy} & = -\rho_{yx} = R_{xyz}H_z \label{r1}\\
	\rho_{yz} & = -\rho_{zy} = R_{yzx}H_x\label{r2}\\
	\rho_{zx} & = -\rho_{xz} = R_{zxy}H_y = R_{yzx}H_y\label{r3}
\end{align}

Since $\rho_{xz}$ shown in Fig. \ref{MR_Hall}(a) is associated with $R_{yzx}$, measurements of either $\rho_{xy}$ or $\rho_{yx}$ are required to fully elucidate the Hall effect of CrSb.
However, measuring the Hall resistivity in the $xy$-plane is difficult due to the thin rod-like shape of the crystals, which is elongated along the $c$ axis ($\parallel z$).
To overcome this issue, we cut an $xy$-oriented lamella out of a single crystal by FIB machining \cite{moll_focused_2018}.

Figure \ref{FIB_RT}(a) presents a schematic drawing of how the lamellae were cut out from the bulk single crystal.
In addition to the $xy$-, a $yz$-oriented lamella was also prepared for comparison.
We refer to these two lamellae as L1 and L2, respectively. 
The extracted lamella was affixed to a substrate that had pre-patterned electrodes, and electrical contacts were formed by FIB-assisted W deposition using $\rm W(CO)_6$ gas as a precursor.
Figure \ref{FIB_RT}(b) shows a scanning ion microscope image of the completed sample with the $xy$-oriented lamella (L1), which was used for the measurements of $\rho_{xx}$ and $\rho_{yx}$.

Figure \ref{FIB_RT}(c) shows the temperature dependence of the normalized resistivity for the three samples. 
A sharp drop in $\rho_{zz}$ (L2) was observed around 6 K with decreasing the temperature.
Because it shifted to lower temperatures by application of magnetic fields as shown in the inset, this is likely a superconducting transition.
It is known that amorphous W including C and Ga impurities deposited from W(CO)$_6$ gas and Ga-source FIB shows a superconducting transition, and the transition temperature is between 5.0 and 6.2 K \cite{sadki_focused-ion-beam-induced_2004, li_tunability_2008}.
We think that such W-based material was unintentionally formed on L2 during the deposition of the W contacts.
In contrast, no superconducting transition was observed in $\rho_{xx}$ (L1), and the temperature dependence was almost identical to that of the bulk sample.

\begin{figure}
	\includegraphics[width=0.8\linewidth]{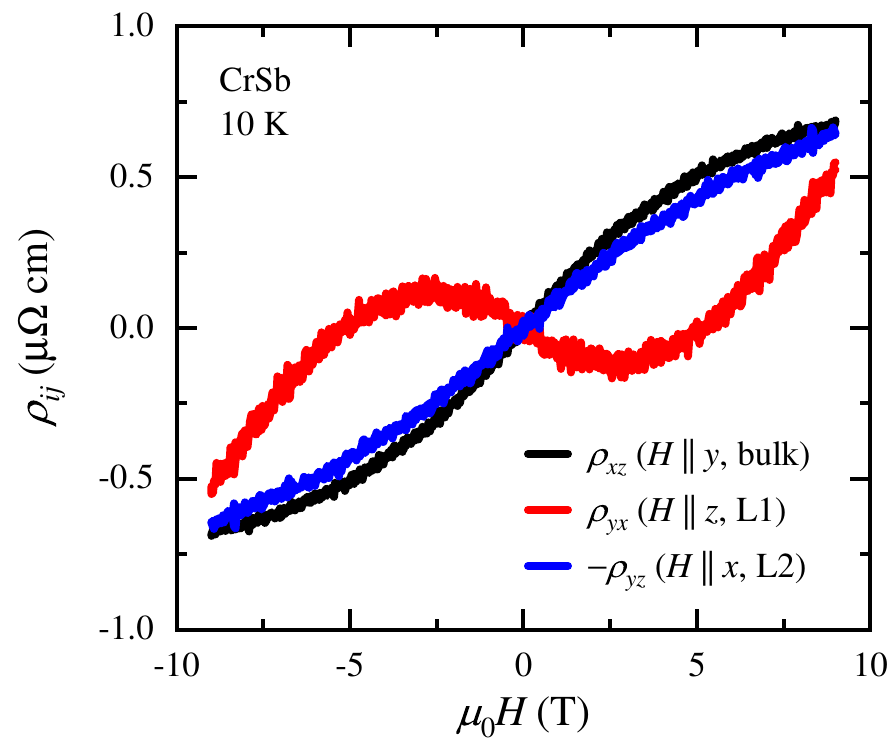}
    \caption{(Color online) 
		    Magnetic field dependence of the three tensor components of Hall resistivity measured at 10 K.
}
\label{10K}
\end{figure}

Figure \ref{10K} shows the magnetic field dependence of the Hall resistivity for the three different configurations.
The measurements were carried out at 10 K, at which the influence of the superconducting transition of the W-based material should be negligible. 
As shown in the figure, all three components exhibited nonlinear field dependence.
$\rho_{xz}$ and $-\rho_{yz}$ were quantitatively very similar as expected from Eqs. (\ref{r2}) and (\ref{r3}), which corroborates that the influence of the W-based material is minimal at this temperature.

\begin{figure}
    \includegraphics[width=1.0\linewidth]{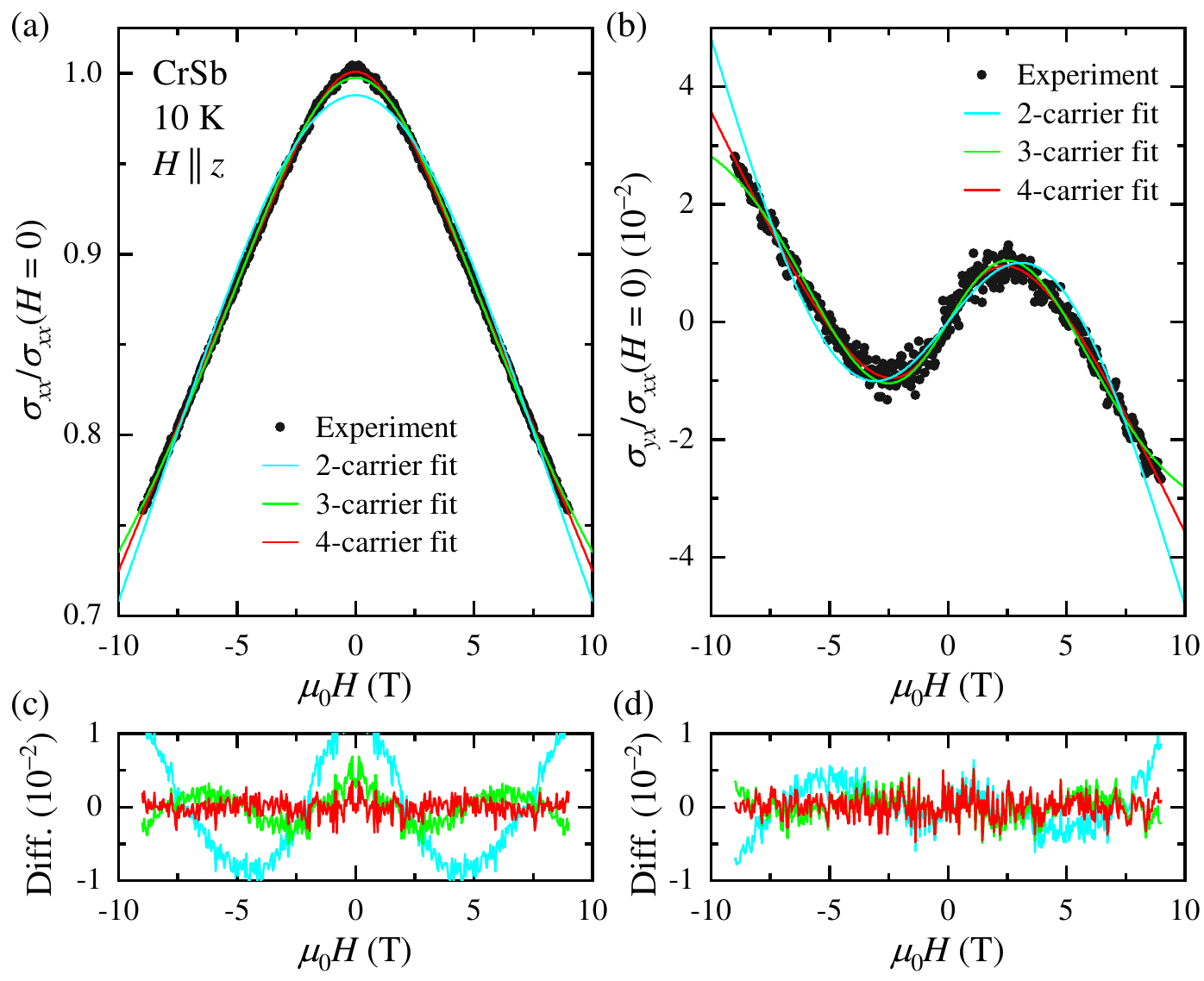}
    \caption{(Color online) Magnetic field dependence of normalized conductivity, (a) $\sigma_{xx}/\sigma_{xx}(H = 0)$ and (b) $\sigma_{yx}/\sigma_{xx}(H = 0)$.
	    The results of multicarrier fitting are also plotted. 
	    The differences between the data and the fits are shown in panel (c) and (d), respectively.
    }
\label{Fitting}
\end{figure}

On the other hand, $\rho_{yx}$ was significantly different from the other two components and displayed a highly nonmonotonic behavior.
The slope of $\rho_{yx}$ changed between negative and positive depending on the magnetic field, suggesting the existence of different types of charge carriers.
More specifically, there should be highly mobile electron-type carriers that dominate the low field behavior and low mobility hole-type carriers that govern the high field region.
To gain more detailed information about the carrier properties, we carried out a multicarrier fitting to the data.
Under the relaxation time approximation, the conductivity tensor components in the plane normal to the applied field can be written as follows \cite{kim1993multicarrier}: 

\begin{equation}
[\sigma_{ij}] = \sum_{n}\frac{eN_n\mu_n}{1 + (\mu_n\mu_0H)^2}
\begin{bmatrix}
1 & s_n\mu_n\mu_0H \\
-s_n\mu_n\mu_0H & 1 \\
\end{bmatrix},
\end{equation}
where the subscript $n$ indicates the carrier index, $e$ the elementary charge, $N_n$ the carrier density, $\mu_n$ the mobility, and $s_n$ takes $-1$ for an electron-type carrier and $+1$ for a hole-type carrier.
These equations were fitted to the conductivity components $\sigma_{xx}$ and $\sigma_{yx}$ with $N_n$ and $\mu_n$ as free parameters.

\begin{table}
  \begin{center}
	\caption{The carrier density $N_n$ and the mobility $\mu_n$ of CrSb deduced from the multicarrier fitting.}
	\label{param}
		\begin{ruledtabular}
    \begin{tabular}{ccc}
         Carrier type & $N_n\, (\rm{cm}^{-3})$ & $\mu_n\, (\mathrm{cm^{2}/V\, s})$ \\
        \hline
	Electron & $2.97\times 10^{20}$ & 640  \\
	Electron & $1.59\times 10^{19}$ & 2400  \\
	Hole & $7.03\times 10^{20}$ & 436  \\
	Hole & $8.73\times 10^{18}$ & 2220  \\
    \end{tabular}
	\end{ruledtabular}
  \end{center}
\end{table}

Figures \ref{Fitting}(a) and \ref{Fitting}(b) show the conductivity components normalized to the zero-field value of $\sigma_{xx}$ and the fitting results.
The fitting was performed by assuming two, three, or four types of carriers. 
The discrepancies between the fits and the data are shown in Figs. \ref{Fitting}(c) and \ref{Fitting}(d).
Systematic differences are evident between the data and the fitting curve when fewer than four types of carriers were assumed, especially for $\sigma_{xx}$ as shown in Fig. \ref{Fitting}(c).
On the other hand, a good fit was obtained by assuming four types of carriers, and the parameters derived from the fitting are shown in table \ref{param}. 
The result indicates the presence of two electron- and two hole-type carriers, with one of each type having low density and notably high mobility.
This is not inconsistent with the existence of the Weyl points that a recent theoretical study had reported to exist in CrSb \cite{wu_magnetic_2023}, because Weyl fermions can exhibit very high mobility owing to the suppression of backscattering \cite{ando_berrys_1998} and a small effective mass.
The coexistence of Weyl fermions and altermagnetism has been claimed for GdAlSi \cite{nag_gdalsi_2023}, where the carrier mobility was $\sim$900\ cm$^2$/Vs \cite{laha_electronic_2024}.
The carrier mobility of CrSb is even higher than this, and to the best of our knowledge, it represents the highest value reported among altermagnets so far.

It is worth mentioning that the high mobility can be an advantage for spin current generation.
It has been shown that collinear antiferromagnets that have anisotropically spin-split bands can generate spin currents and the spin conductivity diverges with suppressing scattering \cite{naka_spin_2019,Hernandez_2021}.
Altermagnets belong to this category of materials.
Hence, the high mobility, together with the recently reported large spin splitting in the band structure near the Fermi level \cite{reimers_direct_2024}, makes CrSb an attractive candidate for an efficient spin-current generator material.

\section{Conclusion}
In conclusion, we found that the Hall resistivity of CrSb exhibited a nonlinear field dependence at low temperatures.
Although it is tempting to attribute this observation to an anomalous Hall effect, symmetry considerations indicate that it is not the case, but more likely to originate from the presence of multiple types of carriers.
A multicarrier analysis revealed the existence of four types of carriers in this system, two of which exhibited mobilities exceeding $2000\ \rm{cm^2/Vs}$.

\section*{Acknowledgment}
The authors are grateful to Hikaru Watanabe for fruitful discussions.
This work was partly supported by Murata Science and Education Foundation.

\bibliography{CrSb}

\end{document}